\documentclass[runningheads]{llncs}

\usepackage{amssymb}
\usepackage{amsmath}
\usepackage{graphicx}
\usepackage{multirow}
\usepackage{url}
\usepackage{subfig}

\DeclareMathOperator{\argmax}{argmax}

\begin{document}

\title{Automatic Segmentation of the Prostate on 3D Trans-rectal Ultrasound Images using Statistical Shape Models and Convolutional Neural Networks}

\author{Golnoosh Samei, Davood Karimi, Claudia Kesch, Septimiu Salcudean}

\vspace{-1mm}


\institute{The University of British Columbia}

\maketitle              

\vspace{-7mm}

\begin{abstract}
In this work we propose to segment the prostate on a challenging dataset of trans-rectal ultrasound (TRUS) images using convolutional neural networks (CNNs) and statistical shape models (SSMs). 
TRUS is commonly used for a number of image-guided interventions on the prostate. Fast and accurate segmentation on the organ in these images is crucial to planning and fusion with other modalities such as magnetic resonance images (MRIs) . However, TRUS has limited soft tissue contrast and signal to noise ratio which makes the task of segmenting the prostate challenging and subject to inter-observer and intra-observer variability. This is especially problematic at the base and apex where the gland boundary is hard to define.  In this paper, we aim to tackle this problem by taking advantage of shape priors learnt on an MR dataset which has higher soft tissue contrast allowing the prostate to be contoured more accurately. We use this shape prior in combination with a prostate tissue probability map computed by a CNN for segmentation. 
\end{abstract}

Keywords: motion compensation, prostatectomy, image registration

\section{Introduction}

TRUS is the medical imaging modality of choice to guide prostate interventions, as TRUS is real-time, easily accessible and the TRUS transducer can be placed proximal to the prostate \cite{samei2020partial,hung2012robotic,samei2018real,golshan2020automatic}. TRUS can image the prostate but cannot accurately image prostate cancer. 
MRI achieves the most accurate prostate cancer imaging; however, it cannot be used during most interventions with existing techniques due to magnetic safety, interference, and limited space. Therefore, in many applications TRUS is used in a fusion process with a pre-operative MRI image. Segmentation of the organ in both images is usually the first step of a surface-based registration process. 

 There has been extensive works on segmenting 2D ultrasound images \cite{ghose2012survey}. For TRUS segmentation, these methods suffer from a lack of image contrast, especially at the prostate base and apex. There have been efforts to use the prostate shape, obtained from MRI, to constrain the prostate segmentation in TRUS \cite{KARIMI2019186,zeng2018,karimi2019accurate}. For segmentation of 3D TRUS images a common approach is to start with the segmentation from the mid-gland and propagate the contour to the neighbouring slices using edge-based or region-based methods  \cite{li2016segmentation,yu2016fully,qiu2015rotationally}. However, these methods are typically much less accurate and less robust than their counterparts for prostate segmentation in MRI \cite{litjens2014evaluation,karimi2019deep,karimi2018prostate}. Also, these methods are not fully automatic. Moreover, propagation-based methods are prone to the accumulation of error from mid-gland to base and apex. The fully-automatic method closest to our work which employs a 3D approach is that of  Zetting et al.~\cite{zettinig2015multimodal}. They used Hough forests to automatically segment the prostate in TRUS images. They report a Dice similarity coefficient (JSC) of 87 on the midglands but do not report the overall error. 
An important prostate segmentation challenge in TRUS images is accuracy at the base and apex.    Due to low signal to noise ratio and tissue contrast between the prostate and its neghbouring tissues such as bladder, a manual segmentation even by an expert which is based solely on the TRUS image without any information on the size, extent along the probe and the shape of the prostate is often unreliable at the base and apex. The use of a ground-truth built in this manner for the purpose of training, leads to sub-par segmentation performance in these regions as reported previously \cite{nouranian2016learning,anas2017clinical}.
To tackle the absence of reliable ground-truth, we propose to use a dataset of paired MR-TRUS images for training.  MR images are used to extract the prostate boundary. An expert will then deform this initial contour using an interactive tool to match it to the surface of  the organ as much as visibility allows. 
 Whereas in segmentation of the prostate on MRI,  machine learning techniques that can learn the texture and local features have been successful at classifying the volume voxels into prostate vs non-prostate \cite{litjens2014evaluation}, this approach is not as successful on TRUS where edges might be missing and texture similarities exist between regions inside and outside of the organ.
 
Statistical shape models (SSM) are powerful tools that are widely used for incorporating top-down information about the expected shape and variation into medical image segmentation methods.  In the last years, the advances in neural networks and their superior performance led to a decline in interest in SSMs for segmentation. However, most recently some important work has been done to incorporate shape priors into the CNNs \cite{KARIMI2019186,zeng2018,karimi2019accurate,milletari2017integrating,ravishankar2017learning}.  While these approaches are promising, more research is still required to fully understand the mechanisms for training CNNs that directly predict shapes. We propose a well studied and intuitive way of leveraging the power of CNNs and SSM.  Instead of relying on CNNs for the final segmentation or shape, we use their classification predictions as a probability map and optimize a utility function that maximizes the probability of the  prostate shape and appearance simultaneously.  In summary, the key contributions of this work are: Using the MRI to compile a reliable ground-truth of the prostate segmentation in TRUS for training and evaluation. 2) Treating the output of the CNN as a probability map instead of a definite segmentation 3) Using the MRI shape model to find the most probable segmentation.

\section{Materials and Methods}
\label{sec:methods}

\subsection{Materials}

We used two  datasets with overlapping patients to validate the proposed method: an MRI  and a TRUS dataset.
Our MR dataset consists of 78 patients. The first 25 were the cases from the MICCAI prostate segmentation challenge (PROMISE12) that did not have an endorectal coil.  
The rest of 53 T2 MRIs came from our  in-house dataset of patients undergoing focal therapy or prostatectomy and were segmented by a collaborating clinical specialist. Each case consists of axial images with 0.27 mm $\times$ 0.27 mm  in-plane resolution and 4 mm slice thickness.  

Our TRUS datasets included a total of 15 patients, 6 prostatectomy (P1-P6) and 9 patients undergoing focal therapy (F1-F9), all of whom had informed consent in institutional review board approved studies. Each patient had between 3 to 9 TRUS images and we had a total of 100 TRUS images.  We used a BK Pro-Focus ultrasound system (BK Medical, Herlev, Denmark) with a biplane TRUS (transverse micro-convex/axial) 8848 4-12 MHz probe.
The TRUS probe was rolled around its axis in $0.9^\circ$ increments ($90^\circ$ in total) to collect $100$ sagittal images of $138\times120$ pixel resolution, with a depth of $60$\,mm and a width of $69$\,mm. These images were interpolated to create a  volume with an isotropic voxel size of $0.5 \times 0.5 \times 0.5$\,mm$^3$.  Note that to record synchronized images with the  motor roll angle, we had to save the raw radiofrequency data (RF), and the processing algorithms we used to generate B-mode images are not comparable to commercial solutions. Moreover, we acquired sagittal slices, which were later interpolated on a 3D grid for the volumes, which thus leads to lower elevational resolution. These factors account for the sub-par quality of the images in our dataset compared to clinical TRUS images.

\subsection{Ground Truth}
Given the difficulty in discerning prostate from the surrounding tissue at the base and apex on TRUS images, the use of MRI segmentation for initialization ensured a better segmentation at these regions. Hence, we used an in-house finite-element based interactive software to manually segment the TRUS images using their MRI. For each patient, the initial segmentation was performed on the respective MRI image. The segmentation contours were overlayed on the TRUS image and the expert first rigidly and then deformably aligned the contours on the TRUS image. For rigid alignment, the expert had control over translation and rotation parameters.   Since we are segmenting the same organ, the deformations could be due to forces applied through the US probe or gravity due to patient's position change and hence of biomechanical nature. We used an interactive graphics interface which models the biomechanical forces applied to the organ boundaries using a finite element method. The expert's input are in forms of displacements to the organ surface. The force is calculated and the resulting displacements for the rest of the organ are applied.

\begin{figure}
  \begin{tabular}{p{.35\linewidth}p{.35\linewidth}p{.35\linewidth}}
\multicolumn{3}{p{1\linewidth}}{\centering \includegraphics[width=1\linewidth]{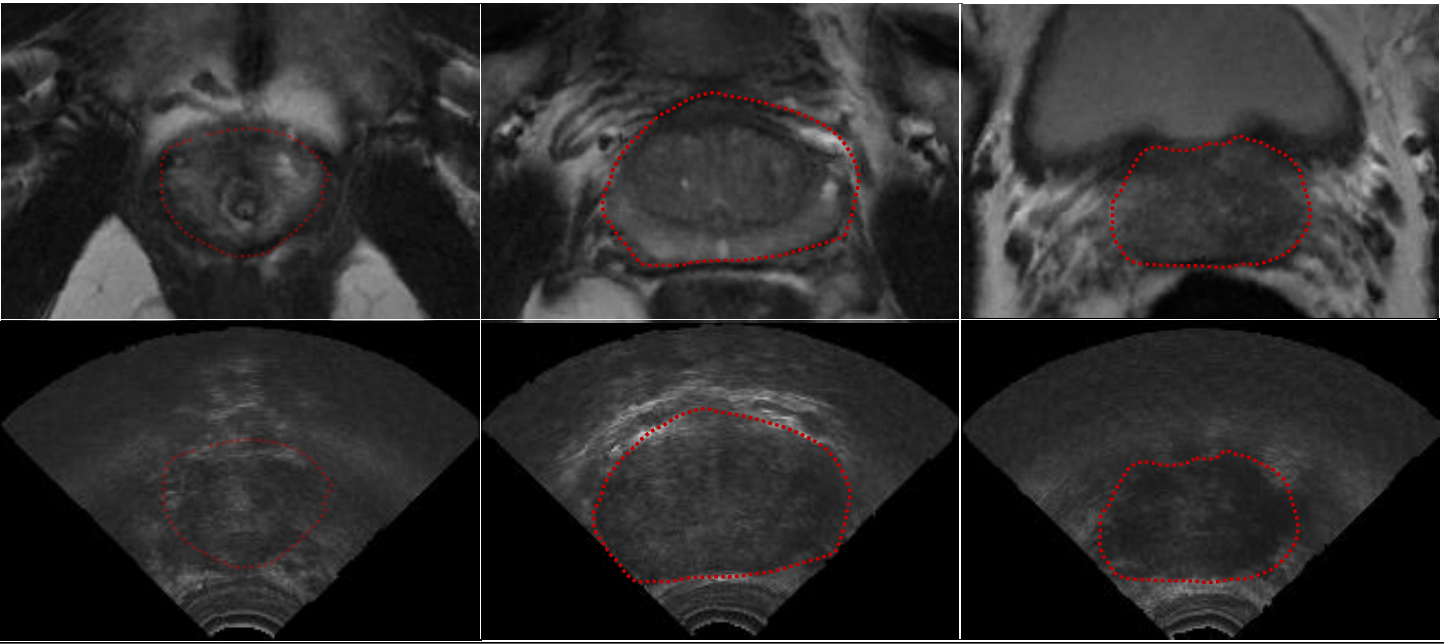}}\\
\multicolumn{1}{p{.33\linewidth}}{\centering(a)} & \multicolumn{1}{p{.33\linewidth}}{\centering(b)}  &
\multicolumn{1}{p{.33\linewidth}}{\centering(c)}\\
\end{tabular}
\caption{An example MRI-TRUS pair in our dataset for generating the ground-truth. Top row displays the registered MR slices with the overlaid segmentation by the clinical expert. Bottom row displays the corresponding TRUS slices with the same segmentation at the (a) apex, (b) midgland and (c) base.  }
\label{fig:3d_2d_images}
\end{figure}
\subsection{CNN architecture}
\label{sec:cnn_architecture}

Our network consists of a contracting path and an expanding path, similar to the V-Net architecture~\cite{milletari2016v}. In brief, in the contracting path we apply successive 3D convolutional filters to compute features at different resolutions. These features will then go through a series of transpose convolutions in the expanding path. Starting with the coarsest feature map, these transpose convolutions aim at recovering the fine detail that is lost in the contracting path. The finer-resolution features computed in the contracting path are concatenated with these features at each step. This will allow the network to directly use the detailed information in the fine convolutional features, which is necessary for accurate delineation of the prostate boundary. Unlike V-Net, however, we apply convolutional filters of different sizes and strides to the input TRUS image. Specifically, we apply convolutional filters of kernel size $(2k+1)$ and stride $k$ for $k \in \{1, 2, 4, 8 \}$ to the image. Our motivation here comes from the lack of edge information in small patches of a TRUS image. Although a succession of several convolutional layers with small kernels will gradually increase the field-of-view, it will still be hard to learn informative features at the first layer because of the lack of any edge information in small patches of the source image. Moreover, in order to increase the richness of the features learned by the network, we will forward the features computed at every fine scale to all succeeding coarser scales as suggested by the DenseNet architecture~\cite{huang2017densely}. For instance, the finest features computed with $k=1$ are forwarded to all coarser layers. All convolution and transpose convolution layers include leaky rectified linear activations. The convolutional feature maps in the last layer of the expanding path will have the same size as the input image. They will go through a final convolution and softmax operation to produce the segmentation probability map in $[0,1]$.

All network weights are initialized randomly using the method proposed in~\cite{he2015delving}. We compute the Dice coefficient of the predicted segmentation with the ground-truth segmentation and the network is trained by minimizing $1-Dice$ using Adam method~\cite{kingma2014adam}. The network was implemented in TensorFlow and trained on an Nvidia Titan V graphics processing unit (GPU). We use a batch size of 1 image due to GPU memory constraints and a learning rate of $10^{-4}$.

\subsection{Statistical Shape Model}

The prostate was manually segmented in our dataset by a clinical expert. Next, the organ contours were resampled into 100 equally distanced points. The prostate surface was represented by these points which are hereafter referred to as the point cloud.
 One of the patient's prostate surface point clouds was set as the reference instance. The other point clouds were all rigidly registered to the reference instance using  the coherent point drift (CPD) algorithm \cite{myronenko2010point} to minimize shape variations due to rotation and translation. Next, the reference point cloud was deformably registered to the rigidly aligned point clouds using non-rigid CPD. This provided the correspondence between the points among the instances.  The registered instances of the reference point cloud consisted of $N=1256$ points and we had $M=76$ instances. Each instance is represented with a $3N$ vector $\mathbf{p} =[x_1,~y_1,~z_1,~~ \cdots ~~,~x_N,~y_N,~z_N]$. We applied PCA to obtain the main modes of variation, $E_i, i=1\cdots M$, of the prostate shape among the population.

\begin{equation} \label{eq:ssm_1}
\mathbf{p} \simeq \mathbf{p}_{\mu} + \sum_{i=1}^{M-1} w_i E_i,
\end{equation}

with $\mathbf{p}_{\mu}$ being the mean shape and $w_i$ being the weight of the $i^{th}$ mode in the summation above.
We noted that the first $C=14$ components accounted for 90\% of the variation of the population and the corresponding reconstruction error, $\|\mathbf{p}-\mathbf{p}_{\mu}-\sum_{i=1}^C w_i E_i\|_2$, was 1.85\,mm.
Hence, each prostate surface is represented using a weight vector $\mathbf{w}=[w_1 ... w_C]$ of size $C=14$ and predicting a segmentation amounts to finding this weight vector in addition to the rigid registration parameters, namely a translation vector $\mathbf{t} \in \mathbb{R}^3$  and rotation parameters $\boldsymbol{\theta} \in \mathbb{R}^{3}$.

\subsection{Shape Fitting and Optimization}

As explained in Sec~\ref{sec:cnn_architecture}, for a TRUS image  $I$ we obtain the prostate probability map, $M(I)$  using our trained CNN. We define the parameter vector $\mathbf{p}=[ \mathbf{w},\mathbf{t},\boldsymbol{\theta}]$ and the resulting prostate binary mask $B(\mathbf{p})$. We find $\mathbf{p}_o = \argmax_{\mathbf{w}} U(I,\mathbf{w})$, an instance of the shape model that best matches the probability map $M(I)$ with the optimum rigid alignments based on the following utility function: 

\begin{equation}
U(I,\mathbf{p}) = \|B(\mathbf{p})\cdot M(I)\|  -\alpha\|\mathbf{w}\|.
\end{equation}

 We used a value of $\alpha=0.1$ throughout our experiments. Initial experiments with gradient based optimizers resulted in convergence problems or high errors.  It is known that
gradient descent methods are sensitive to initialization, and are not suitable for our nonconvex optimization problem.  Therefore, we employed particle swarm optimization
introduced by \cite{ebenhart4kennedy}, which belongs to the category of population-based optimization algorithms and are known for overcoming such problems.
\section{Experiments and results}
We performed leave-one-out experiments. For each patient, we built a separate shape model based on all the images in our MRI dataset excluding its own MRI. 
Also we separately trained a CNN on all the TRUS images excluding  those of the left-out patient.  We used the Jaccard similarity coefficient (JSC) to evaluate our method. The results are presented in Table. \ref{tab:results}. 

\begin{table}[t]
\caption{Statistics\,(\,mean\,$\pm$\,std) of the JSC over the patients at three different regions and overall.}
\label{tab:results}
 \centering
        \small
        \setlength\tabcolsep{2pt}
\begin{tabular}[t]{|c|c c|c c |c c |c c |}
  \hline
Patients &  \multicolumn{2}{|c|}{Overall} & \multicolumn{2}{|c|}{Apex}  & \multicolumn{2}{|c|}{Mid-gland}  & \multicolumn{2}{|c|}{Base}   \\
\hline
\hline
 F1  & 92.46 $\pm$ &6.73 & 91.63 $\pm$ &10.20 & 92.60 $\pm$ &5.31 & 87.45 $\pm$ &5.11\\
 F2  & 90.42 $\pm$ &0.61 & 92.58 $\pm$ &3.98 & 93.10 $\pm$ &1.65 & 91.59 $\pm$ &1.97\\
 F3  & 88.85 $\pm$ &1.14 & 90.47 $\pm$ &6.42 & 93.05 $\pm$ &1.62 & 88.12 $\pm$ &2.43\\
 F4  & 90.04 $\pm$ &1.61 & 91.34 $\pm$ &3.62 & 94.89 $\pm$ &1.14 & 90.27 $\pm$ &1.46\\
 F5  & 90.41 $\pm$ &4.39 & 93.68 $\pm$ &6.68 & 95.01 $\pm$ &7.60 & 87.90 $\pm$ &1.33\\
 F6  & 90.30 $\pm$ &1.51 & 91.75 $\pm$ &7.63 & 95.01 $\pm$ &2.37 & 87.90 $\pm$ &9.74\\
 F7  & 88.67 $\pm$ &3.73 & 92.64 $\pm$ &4.54 & 96.25 $\pm$ &1.92 & 92.37 $\pm$ &1.56\\
 F8  & 89.54 $\pm$ &1.26 & 90.75 $\pm$ &8.18 & 95.54 $\pm$ &0.94 & 89.07 $\pm$ &1.35\\
 F9  & 91.50 $\pm$ &2.63 & 89.79 $\pm$ &9.91 & 95.01 $\pm$ &1.00 & 89.97 $\pm$ &1.63\\
\hline
 P1  & 88.07 $\pm$ &2.72 & 92.26 $\pm$ &1.64 & 92.59 $\pm$ &0.60 & 89.00 $\pm$ &6.34\\
 P2  & 90.74 $\pm$ &0.53 & 93.26 $\pm$ &2.09 & 95.19 $\pm$ &0.58 & 91.33 $\pm$ &0.99\\
 P3  & 92.32 $\pm$ &3.58 & 91.47 $\pm$ &4.00 & 92.60 $\pm$ &1.00 & 87.37 $\pm$ &3.01\\
 P4  & 89.32 $\pm$ &1.75 & 88.19 $\pm$ &5.81 & 95.89 $\pm$ &1.26 & 88.76 $\pm$ &2.23\\
 P5  & 92.80 $\pm$ &3.25 & 91.59 $\pm$ &1.12 & 96.13 $\pm$ &4.84 & 93.49 $\pm$ &3.06\\
 P6  & 91.88 $\pm$ &2.90 & 93.75 $\pm$ &8.81 & 94.86 $\pm$ &2.34 & 92.80 $\pm$ &2.13\\
\hline
total& 90.58 $\pm$ &4.69 & 91.50 $\pm$ &8.98 & 94.51 $\pm$ &5.34 & 89.83 $\pm$ &4.23\\
\hline
\end{tabular}
\end{table}

\begin{figure}
  \begin{tabular}{p{.25\linewidth}p{.25\linewidth}p{.25\linewidth}p{.25\linewidth}}
\multicolumn{3}{p{1\linewidth}}{\centering \includegraphics[width=1\linewidth]{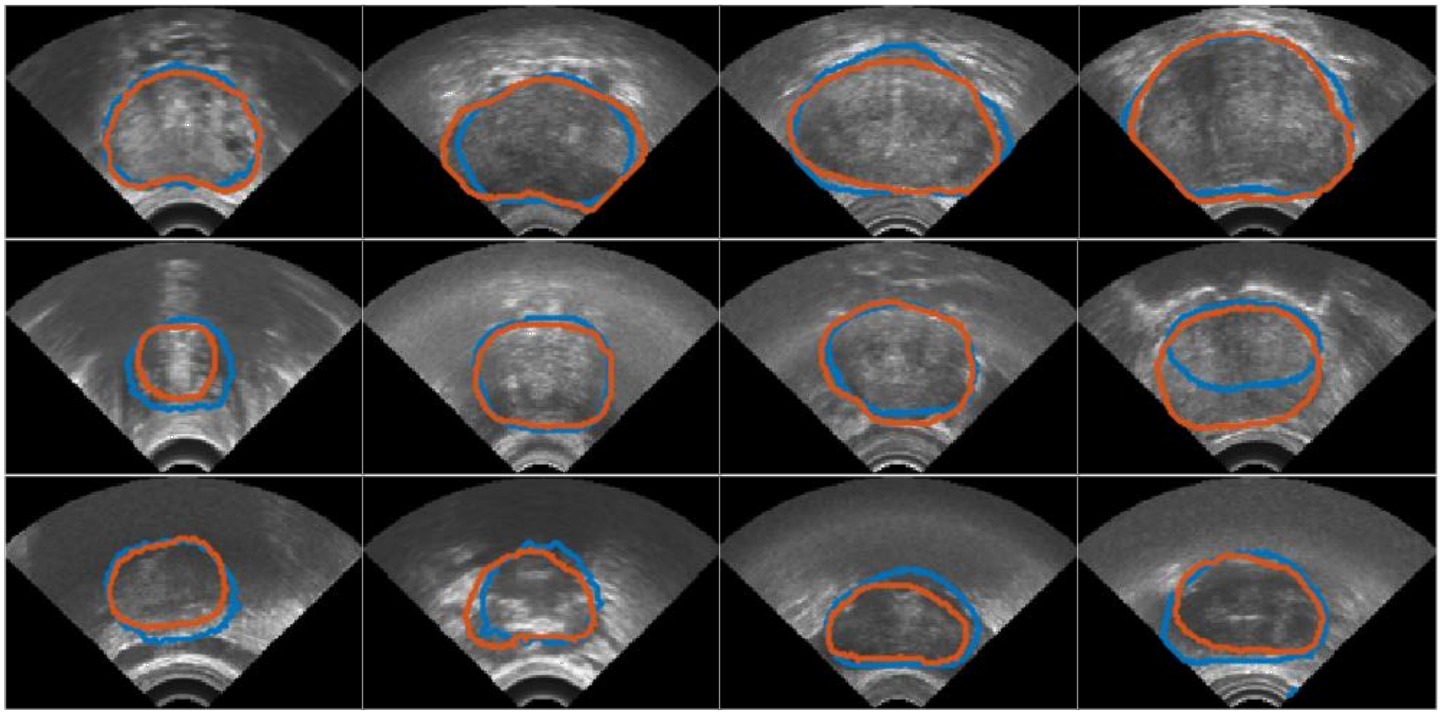}}\\
\end{tabular}
\caption{Example of results on four different patients. Top row is a slice in midgland, middle row a slice on apex and bottom row a slice on base.  The ground-truth segmentations is displayed with red and our segmentation is displayed in blue.}
\label{fig:results}
\end{figure}
\section{Discussion and Conclusion}
\label{sec:conclusion}
We have presented a method to segment the prostate on a challenging dataset of TRUS images. Our method has a number of novelties. Firstly, we are tackling the known problem of unclear boundaries at the base and apex using a paired MR-TRUS dataset. We use the MR images in two ways, firstly we use the organ segmentation in an MRI of the same patient to guide the manual segmentation.   Secondly, we use the MRI dataset to build a statistical shape model. 
Training the CNN based on this ground-truth leads to the network learning the prostate boundaries based on other visible structures. Hence we believe our CNN assigns high probabilities to regions at the base and apex based on neighbouring structures such as bladder or seminal vesicles. This has been learned through the training process based on training examples. 

Our second novelty is to define a utility function based on the probability maps generated by CNN and the likelihood of a given shape based on SSM. This helps recover parts of the organ that are not present in the TRUS image and or constrain the results to statistically plausible results. 

Once we have trained our CNN and SSM, our method can be employed to perform segmentation for a new patient, without the need to an MRI. 
We evaluated our method using the leave-one-out scheme on a dataset of 100 3D TRUS images.  Our results are comparable to the state-of-the art at the mid-gland and superior at the apex.

\bibliographystyle{plain}
\bibliography{miccai2018}   
\end{document}